\documentclass[pra,superscriptaddress,a4paper,twocolumn,nofootinbib,showpacs]{revtex4}
\usepackage{amsmath,amsfonts,graphicx,color,bbm}
\usepackage{algorithmic}

\def\>{\rangle}
\def\<{\langle}
\begin{document}

\title{Transitions in the computational power of thermal states for measurement-based quantum computation} %
\author{Sean D. Barrett}%
\affiliation{Optics Section, Blackett Laboratory, Imperial College
London, London SW7 2BZ, United Kingdom}%
\author{Stephen D. Bartlett}%
\affiliation{School of Physics, The University of Sydney,
  Sydney, New South Wales 2006, Australia}%
\author{Andrew C. Doherty}%
\affiliation{School of Physical Sciences, The University of
  Queensland, St Lucia,
  Queensland 4072, Australia}%
\author{David Jennings}%
\affiliation{School of Physics, The University of Sydney,
  Sydney, New South Wales 2006, Australia}%
\author{Terry Rudolph}%
\affiliation{Optics Section, Blackett Laboratory, Imperial College
London, London SW7 2BZ, United Kingdom}%
\affiliation{Institute for Mathematical Sciences, Imperial College
  London, London SW7 2BW,
United Kingdom}%

\date{18 December 2009}

\begin{abstract}
  We show that the usefulness of the thermal state of a specific spin-lattice model for measurement-based quantum computing exhibits a transition between two distinct ``phases'' -- one in which every state is a universal resource for quantum computation, and another in which any local measurement sequence can be simulated efficiently on a classical computer.  Remarkably, this transition in computational power does \emph{not} coincide with any phase transition, classical or quantum, in the underlying spin-lattice model.
\end{abstract}

\maketitle

\section{Introduction}

Determining whether properties of the ground or thermal state of a
coupled quantum many-body system can be efficiently simulated on a
computer is a central question in many-body physics. Consider a spin lattice model that is unitarily equivalent to a lattice of uncoupled spins; in such a model, the spectrum and the behaviour of simple correlation functions can be solved analytically.  However, such simplicity can belie an underlying complexity, as the calculation of more general observables on the thermal state of this system can be computationally intractable. For example, one might want to simulate the outcomes of a sequence of local (single-spin) measurements on a subset of spins; such measurements are used to reveal ``hidden'' order such as long-range entanglement~\cite{Pop05}.  Although numerical techniques such as Monte Carlo methods~\cite{Pop05} may be efficient in some instances, one can also devise models for which the outcomes of such measurements performs a quantum computation~\cite{Rau01,Rau03} for which no classical simulation algorithm is expected to be efficient.  Little is known about how to characterize the classical simulation complexity of such general observables, even in the simplest models.

Here, we provide a remarkable example of a solvable model wherein the efficiency of simulating general observables of the thermal state undergoes a transition, separating the model's parameter space into two distinct regions. In one region, at high temperatures, we present an explicit algorithm that can efficiently simulate the outcome of any local adaptive measurement sequence. In the other region, we prove (based on standard complexity assumptions\footnote{It is believed that local measurements on such states are impossible to efficiently simulate on a classical computer, since, if this were possible, any polynomial time quantum algorithm could be performed efficiently on a classical device.}) that no such efficient simulation algorithm exists.  Remarkably, there is \emph{no} phase transition in this model, classical or quantum, that could serve to demarcate these two regions.

Specifically, we make use of the \emph{cluster state} model of measurement-based quantum computation (MBQC)~\cite{Rau01,Rau03},
wherein a fixed resource state is subjected to local measurements~\cite{Gro07}.  We consider a spin-lattice model for which the unperturbed ground state is a cluster state but which is subject to both thermal noise and a uniform external field. We introduce a method for efficient classical simulation of local measurements on sufficiently high-temperature equilibrium states of this model. Our method significantly generalizes the separability result for bipartite partitions of thermal cluster  states~\cite{Rau05} by combining ideas from mixed-state entanglement and percolation theory. For sufficiently low temperatures and small external fields, we describe a method for obtaining cluster states on which MBQC can be performed by making use of the error-correcting thresholds of~\cite{Rau06,Rau07}.

Our paper is structured as follows.  In Sec.~\ref{sec:Model}, we present our completely-solvable model.  We provide a lower bound on the temperature for which MBQC on the thermal state can be efficiently classically simulated in Sec.~\ref{sec:RegionC}, and an upper bound on the temperature for which the thermal state is a universal resource for MBQC in Sec.~\ref{sec:RegionQ}.  Conclusions are presented in Sec.~\ref{sec:Conclusions}.

\section{Cluster-state model}
\label{sec:Model}

The model we consider is the cluster state Hamiltonian~\cite{Rau05}
with the addition of a local magnetic field in the $z$-direction. A cluster state is a highly entangled state of two-level systems on a lattice $\mathcal{L}$~\cite{Rau01,Rau03}. It can be characterized as the unique $+1$ eigenstate of a set of commuting operators $K_i =
X_i{\textstyle \prod_{j\sim i}}Z_j$, where $X_i$ ($Z_i$) is the Pauli $X$ ($Z$) operator at site $i$ and where $j{\sim}i$ denotes that $j$ is connected to $i$ by a bond in the lattice $\mathcal{L}$.  The cluster state Hamiltonian~\cite{Rau05,Bar06} is a model of interacting spins with $H = - \tfrac{\Delta}{2}\sum_{i \in \mathcal{L}} K_i$ for which the cluster state is the unique ground state, and which possesses an energy gap of $\Delta$ between the ground and first excited states. We will consider a modified Hamiltonian with a local $Z$ field at each site:
\begin{equation}
  \label{eq:ClusterLocalZ}
  H_\theta = - \tfrac{\Delta}{2} {\textstyle\sum}_{i \in \mathcal{L}} (\cos\theta\, K_i + \sin\theta\, Z_i) \,,
\end{equation}
for $0\leq \theta \leq \pi/2$.  With this parameterization, the ground state energy gap of the Hamiltonian is $\Delta$ for all values of $\theta$.  The parameter $\theta$ quantifies the relative strength of the local magnetic field term.  For $\theta=0$ we recover the cluster state Hamiltonian, while for $\theta=\pi/2$ the spins are uncoupled and the ground state is a product state.

The thermal state of this model can be found straightforwardly for any strength of external field and temperature $T$ because the system is unitarily equivalent to a lattice of uncoupled spins~\cite{Ple07}, as follows.  The controlled-phase gate $\text{CZ}=\text{diag}(1,1,1,-1)$ can be applied to any bond in the lattice; because of its symmetry, it does not depend on which qubit is the control and which is the target.  Also, as the gate is diagonal in the $Z$-basis, its action on different bonds all commute with each other.  Therefore, we can define the unitary $\text{CZ}_{\mathcal{L}}$ which is given by the product of controlled-phase gates on all bonds of the lattice.  This unitary operator decouples the system, as
\begin{align}
  (\text{CZ}_{\mathcal{L}})Z_i (\text{CZ}_{\mathcal{L}}^\dagger)&=Z_i \, , \\
(\text{CZ}_{\mathcal{L}})K_i (\text{CZ}_{\mathcal{L}}^\dagger)&=X_i\,,
\end{align}
for all $i \in \mathcal{L}$.  We can therefore express the transformed Hamiltonian as
\begin{equation}
(\text{CZ}_{\mathcal{L}})H_\theta (\text{CZ}_{\mathcal{L}}^\dagger) = - \tfrac{\Delta}{2} {\textstyle\sum}_{i \in \mathcal{L}} (\cos\theta\, X_i + \sin\theta\, Z_i) \,,
\end{equation}
i.e., as a sum over single spin observables. Note that $\text{CZ}_{\mathcal{L}}$ is a non-local unitary that creates entanglement, so that the ground state of $H_\theta$ can be highly entangled even though $H_\theta$ is unitarily equivalent to a model of uncoupled spins.  Note that this solution does not imply that MBQC can be efficiently simulated for any value of $\theta$ or $T$, because the statistics of measurement outcomes may still be hard to calculate.

%% Define the two qubit unitary operator $\text{CZ}=diag(1,1,1,-1)$, and
%% by $\text{CZ}_{\mathcal{L}}$ denote this operator acting on every pair
%% of qubits joined by a bond in the lattice.

This solution yields an explicit expression for the thermal states of (\ref{eq:ClusterLocalZ}). Define the single spin states
\begin{equation}
|\theta\rangle \equiv \cos(\theta/2)|+\>+\sin(\theta/2)|-\>\,,
\end{equation}
with $|\pm\rangle$ the eigenstates of $X$.  The ground state of
(\ref{eq:ClusterLocalZ}) is obtained by applying
$\text{CZ}_{\mathcal{L}}$ to the product state $|\theta
\rangle^{\otimes \mathcal{L}}$ where each spin on $\mathcal{L}$ is
prepared in the state $|\theta\rangle$. We recognize the
ground state of our model as one of the states whose usefulness for
MBQC was considered in~\cite{vdN06}. The thermal state
$\rho(\beta,\theta)$ is obtained by applying $\text{CZ}_{\mathcal{L}}$ to the thermal states
\begin{equation}
\tfrac{1}{2}(I + \tanh(\beta \Delta
/2)e^{i\theta Y_i/2} X_i e^{-i\theta Y_i/2})\,,
\end{equation}
of each independent spin, where $\beta = (k_BT)^{-1}$.

The model is unitarily equivalent to uncoupled spins, and so there are no phase transitions, quantum or classical. Specifically, in the thermodynamic limit, the free energy per site is analytic for the full range of $T$ and $\theta$. Despite this, we now demonstrate that this system exhibits a transition in its usefulness as a resource for universal MBQC.  First, we demonstrate that above a certain finite temperature it is possible to efficiently classically simulate any attempt at MBQC. We label this region of the parameter space of the model `Region $C$'. Second, we demonstrate that for $\theta$ and $T$ below some specific thresholds, the thermal state is a universal resource for MBQC; this is denoted `Region $Q$'.

\section{Region $C$: efficiently simulatable}
\label{sec:RegionC}

We define classical simulation of an attempted MBQC as
in Ref.~\cite{Bra07}: Let $N$ be the number of two-level systems in a particular finite sized lattice.  An MQBC on these qubits can be efficiently classically simulated if there exists a randomized algorithm that can sample results of arbitrary single-qubit measurements (together with feed-forward) from the correct (quantum-mechanical) probability distribution using resources that scale polynomially in $N$ on a classical computer.

\subsection{Simulation of thermal states}

Certain Monte Carlo algorithms simulate properties of a thermal density matrix $\rho(\beta,\theta)$ by sampling energy eigenstates with probabilities given by the Gibbs distribution. Here we will develop a stochastic simulation that efficiently samples from a different set of states $|C_\alpha\>\<C_\alpha|$ with probabilities $p_\alpha$ such that
\begin{equation}
  \rho(\beta,\theta)=\sum_\alpha p_\alpha |C_\alpha\>\<C_\alpha|\,.
\end{equation}
(We note that the states $|C_\alpha\rangle$ need not be orthogonal, and do not in general form a basis.)  The states $|C_\alpha\>$ will be be chosen such that, for sufficiently high temperatures, they have such small amounts of entanglement that they can be stored and manipulated efficiently on a classical computer. We conclude that it is not possible to perform MBQC on the states $\rho(\beta,\theta)$ in the regime of temperature in which this simulation succeeds.

\begin{figure}
  \includegraphics[width=3.25in]{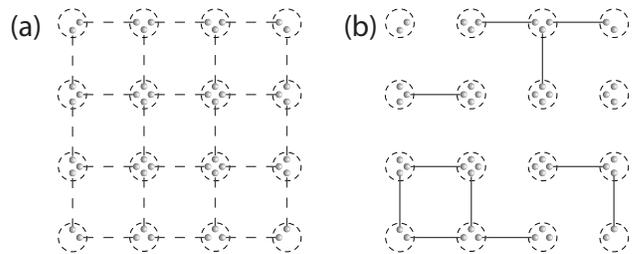}
\caption{(a) A PEPS representation of the thermal state on a square
  lattice. Bonds between virtual qubits, denoted by dashed lines, are
  in a mixed state.  (b) An instance of this thermal state.  Solid
  lines denote a maximally-entangled state on the bond, whereas no
  line denotes a separable state. }
 \label{fig:PEPSpercolation}
\end{figure}

The first step in our simulation algorithm is to represent the state $\rho(\beta,\theta)$ as a \emph{projected entangled pair state}
(PEPS)~\cite{Ver04}; a class of states that generalizes the well-known valence bond solids~\cite{Aff87}. In a PEPS representation, as indicated in Figure~\ref{fig:PEPSpercolation}(a) for a square lattice, a set of $d(i)$ virtual quantum systems (we restrict ourselves to considering virtual qubits) is associated with each site $i$ of the lattice, where $d(i)$ is the number of bonds emanating from site $i$. For every pair of sites $i$ and $j$ connected by a bond, we can identify two virtual qubits (one at site $i$ and one at site $j$) which we associate to this bond. A PEPS on the original lattice is obtained by placing every such bond in some state $\rho_{\text{bond}}$, and at every site $i$ some operator $A_i$ maps the $d(i)$ virtual systems onto the physical system. (The operators $A_i$ are isometries, but are commonly referred to as projectors.)

For the cluster state (and generally for PEPS representations), the
pair of virtual qubits associated to each bond are assigned the
maximally-entangled state $\rho_{\text{bond}} = |C_2\rangle\langle C_2|$, with
\begin{equation}
|C_2\rangle=\tfrac{1}{\sqrt{2}}\bigl(|0\rangle|+\rangle+|1\rangle|-\rangle \bigr)\,,
\end{equation}
where $|\pm \rangle=\frac{1}{\sqrt{2}}(|0\rangle \pm |1\rangle)$.
With this convention, the cluster state has a simple PEPS
representation~\cite{Ver04} corresponding to the choice of isometry
\begin{equation}
  A=|0\rangle \langle00\ldots 0|+|1\rangle \langle11\ldots 1|\,,
\end{equation}
at each site, with $d(i)$ zeros (ones) in $\langle00\ldots 0|$
($\langle11\ldots 1|$), and the states $|0\rangle$ and $|1\rangle$
forming a basis for the resulting qubit at each site.

In our simulation algorithm, we follow~\cite{Rau05} and choose more
general states for the pairs of virtual qubits on a bond, allowing
us to describe thermal states of our cluster Hamiltonian, both with
and without the local field term.
The zero-field thermal state $\rho(\beta,0)$ is
obtained as a PEPS if we choose
\begin{equation}
  \label{eq:VBSthermal}
  \rho_{\text{bond}}(\beta,\theta=0) =
  \tfrac{1}{4} (I + \omega_e X \otimes Z) (I + \omega_e Z \otimes
  X)\,,
\end{equation}
on every bond, and
\begin{equation}
  A=|0\rangle \langle 00\ldots 0| + |1\rangle\langle 11\ldots 1|\,,
\end{equation}
at every site~\cite{Rau05}.  For $\omega_e=1$, the bond states are pure and maximally entangled, and the resulting PEPS is the cluster state.  For $\omega_e=0$, the bond states are maximally mixed, corresponding to infinite temperature. To obtain $\rho(\beta,0)$, the parameters $\omega_e$ are chosen such that
\begin{equation}
  \prod_e \omega_e=\tanh(\beta \Delta / 2)\,,
\end{equation}
where the product is taken
over the bonds emanating from a particular site; for simplicity, we
choose all $\omega_e$ to be equal. The case of non-zero $\theta$ is
handled by a slightly more general ansatz for the bond states, as follows
\begin{equation}
  \label{eq:VBSthermal2}
\rho_{\text{bond}}(\beta,\theta) =
\tfrac{1}{4} (I + \alpha X \otimes Z+ \gamma Z\otimes I) (I + \alpha Z \otimes X + \gamma I\otimes Z)\,.
\end{equation}
The details of how to choose $\alpha$ and $\gamma$ to reproduce
$\rho(\beta,\theta)$ is described in Appendix~\ref{sec:AppB}.

The second step of the algorithm is to decompose the states
$\rho_{\text{bond}}$ as an ensemble of states, such that the elements of the ensemble are product states with high probability.  Every two-qubit state has a decomposition of the form~\cite{Lew98}
\begin{equation}
  \rho_{\text{bond}} =  p_e |\psi\rangle\langle\psi | + {\textstyle
  \sum_\mu} p_{s,\mu} |\phi_{1,\mu}\rangle\langle\phi_{1,\mu}| \otimes
  |\phi_{2,\mu}\rangle\langle\phi_{2,\mu}| \,,
  \label{eqn:BSA}
\end{equation}
where $|\psi\rangle$ is a pure entangled state, $|\phi_{1,\mu}\rangle \otimes |\phi_{2,\mu}\rangle$ is a pure product state, $p_e$, $p_{s,\mu}$ are probabilities, and the probability of getting a product state $\sum_\mu p_{s,\mu} \equiv 1-p_e$ is maximal over all possible decompositions.  Following~\cite{Wel01}, we find that for $\rho_{\text{bond}}(\beta,0)$ the state $|\psi\rangle$ is simply the two-qubit cluster state, and that
\begin{equation}
  p_e = ( \omega_e^2 + 2 \omega_e -1 )/2\,.
  \label{eqn:pe}
\end{equation}
As expected, high temperatures yield small values for $p_e$ and an
ensemble that is largely made up of product states. For the case of
general $\theta$ this decomposition is readily performed numerically.

Our key insight for simulating MBQC on a thermal PEPS state is that, armed with a decomposition such as Eq.~(\ref{eqn:BSA}), one can
efficiently generate instances of an ensemble that gives the state
$\rho(\beta,\theta)$.  If the temperature is sufficiently high, the
members of this ensemble can be efficiently stored and manipulated in a classical computer. We outline the simulation algorithm here; full details are given in Appendix~\ref{app:Sampling}.

On a given run of the simulation, the state for each bond is chosen to be either the pure two-qubit entangled state $|\psi\rangle$ or one of the product states $|\phi_{1,\mu}\rangle \otimes
|\phi_{2,\mu}\rangle$.  We note that the operator $A$ can be viewed as one outcome of a generalized quantum measurement, albeit one that reduces the dimension of the measured system from four spins to one. The probabilities for these states are given by the decomposition
(\ref{eqn:BSA}), conditional on performing this measurement and
successfully obtaining the outcome associated with $A$.
Fig.~\ref{fig:PEPSpercolation} illustrates this sampling. For
sufficiently small $p_e$, percolation theory tells us that the lattice decomposes into disconnected clusters with entanglement only within a cluster, and not between clusters. Specifically, if $p_e$ is smaller than the critical bond percolation probability
$p_{\text{crit}}^{\text{bond}}$ for the lattice, then we expect all the clusters will be sufficiently small to store and simulate efficiently on a classical computer.

\subsection{Classical resource requirements}
\label{app:Cluster}

We now provide an explicit bound on the classical computing resources required to execute our simulation algorithm.  Specifically, we prove that if $p_e <
p_{\text{crit}}^{\text{bond}}$, then the total simulation cost is
bounded by a polynomial in $N$.

In our simulation, each run yields a lattice with some entangled and
some product-state bonds.  We identify regions -- ``clusters'' --
that are connected by entangled bonds.  Distinct clusters are
therefore in a product state, and can be stored and simulated
separately. For cluster $C_j$ of size $|C_j|$, a direct simulation
would have a cost proportional to $2^{2|C_j|}$. Because there are at
most $N$ clusters, the total cost of classically simulating a
measurement sequence on the state $\rho(\beta,\theta)$ is bounded
above by $N2^{2|C_j|}$ for each round of the simulation.

We expect that for sufficiently high temperatures $p_e$ will become
very small and the resulting clusters will be small enough to
simulate efficiently. To confirm this expectation, we invoke results
from percolation theory~\cite{Gri89,Bro08} to bound $|C_j|$.  The
essential idea is that, if $p_e$ is smaller than the critical bond
percolation probability $p_{\text{crit}}^{\text{bond}}$ for the
lattice, then all the clusters will be ``small enough'' to simulate
efficiently. Specifically, if $p_e<p_\text{crit}^{\text{bond}}$,
then the mean cluster size $\chi(p_e)$ does not depend on $N$.  In
addition, the largest cluster size is almost surely of size
$\mathcal{O}(\log_2 N)$ with standard deviation $\mathcal{O}(\log_2
\log_2 N)$~\cite{Baz00}.
%, and moreover the probability that a given cluster $C$
%has more than $n
%>\chi(p)^2$ sites is bounded as~\cite{Gri89}
%\begin{equation}
%  \Pr\left[|C|\geq n\right] \leq 2\exp(-n/2\chi^2)\,.
%\end{equation}
Now imagine we reserve $N$ classical registers, each of $kN^c$
classical bits for some constants $k$ and $c$, and use each register
to store the state of one of the $M$ ($\leq N$) distinct clusters.
This allows us to store, with $k$ bit precision, the quantum state
of any cluster for which $|C_j|\le \log_2N^c$, and to simulate local
measurements on it efficiently.  As the largest cluster is almost
surely of this size,
%The probability that we can store \emph{all} of the clusters in this
%memory is given by the probability that \emph{none} of the clusters
%exceeds $\log_2N^c$ qubits in size, and this probability satisfies
%\begin{align}
%  \Pr[|C_j|\leq \log_2 N^c \, ,\forall j] & \geq \bigl(1-\Pr[|C|\geq \log_2 N^c]\bigr)^{M} \nonumber\\
%  & \geq \bigl(1-2\exp(-\tfrac{\log_2 N^c}{2\chi^2})\bigr)^N\,.
%  \label{eqn:SuccessProbabilityBound}
%\end{align}
%If we require that the simulation succeeds with probability
%\begin{equation}
%  \Pr[|C_j|\leq \log_2 N^c \, ,\forall j] \geq 1-\delta\,,,
%\end{equation}
%for some fixed $\delta > 0$, from
%Eq.~(\ref{eqn:SuccessProbabilityBound}) we find that it is
%sufficient to have
%\begin{equation}
%  c\ge 2\chi^2 \ln 2 ((\log_2 N)^{-1}-\log_2(1-\sqrt{1-\delta}))\,,
%\end{equation}
the total simulation cost is bounded by a polynomial in $N$. Note
that while this analysis shows efficient classical simulatability is possible, it is certainly not optimal -- more precise estimates
would require considering the distribution of cluster sizes.

\subsection{Critical temperature for region $C$}

\begin{figure}
  \includegraphics[width=3.25in]{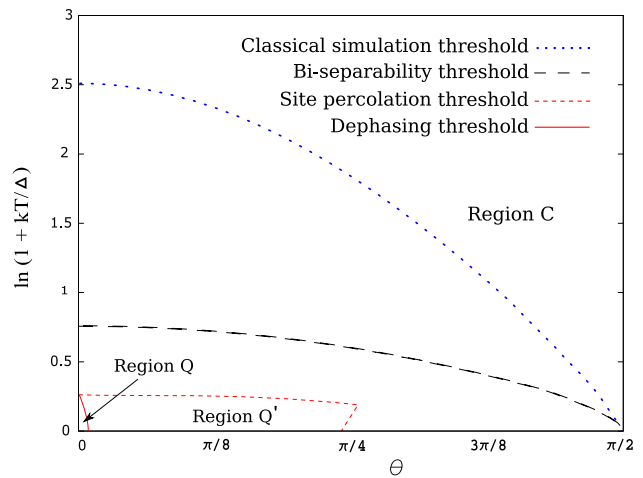}
\caption{The parameter space of our model, with bounds on the usefulness of the thermal state on a simple cubic lattice for MBQC.  Region C represents the region on which any MBQC scheme performed on the thermal state can be efficiently simulated on a classical computer.  Region Q represents the region where we can rigorously show, via the dephasing channel argument given in the text, that the state is universal for MBQC using ideal measurements. Region Q' is a region where the local filtering yields a percolated thermal cluster state with $T<0.28\Delta$; it is possible to perform universal MBQC on any ground state or $\theta{=}0$ thermal state in this phase, and states in this phase with $T,\theta>0$ may be universal for MBQC although rigorous error correction thresholds for such states are not currently known. The black dashed curve represents the separability criteria of~\cite{Rau05}; above this curve, the PEPS description of the thermal state is separable along any bipartite division of the cubic lattice given by a plane.}
 \label{fig:PhaseDiagram}
\end{figure}

With this analysis, we can now lower bound the critical temperature
above which our simulation algorithm succeeds. In the case of
$\theta=0$, we can first compute the critical value of
$\omega_e\equiv\omega_{\mathrm{crit}}$, and relate this to a critical temperature via
\begin{equation}
  \tanh(\beta_\text{crit} \Delta / 2) = \prod_e
  \omega_e = \omega_{\mathrm{crit}}^d\,.
\end{equation}
For some well known lattices we find that $kT_\text{crit}$ is 0.813$\Delta$ (Honeycomb), 1.6921$\Delta$ (Square), 7.1617$\Delta$ (Triangular), 13.1$\Delta$ (Cubic). For $\theta > 0$, we resort to numerical methods (see Appendix~\ref{sec:AppB}) and again find a critical temperature.  As expected, this temperature becomes zero at $\theta = \pi/2$, when the ground state is a product state. Figure~\ref{fig:PhaseDiagram} plots this solution for a cubic lattice.  We note that improved lower bounds of this critical    temperature can be obtained by allowing the value of $\omega_e$ to  vary from bond to bond.

\section{Region $Q$: universal for MBQC}
\label{sec:RegionQ}

We now demonstrate that, for the `cold and weak field' region in parameter space, the thermal state of (\ref{eq:ClusterLocalZ}) is a universal resource for MBQC on an appropriate lattice.  Specifically, we use the error thresholds of~\cite{Rau06,Rau07}, together with a local filtering method, to prove the existence of a finite region of parameter space for which the thermal state of (\ref{eq:ClusterLocalZ}) is a universal resource for MBQC given ideal single-qubit measurements.

Consider performing the local measurement on every site in
$\mathcal{L}$ as introduced in~\cite{vdN06} for the zero temperature case, described by the measurement operators %~\cite{nielsen2000a}
\begin{equation}
  \label{eq:Procrustean}
M_0 = \sqrt{1-\tan^2\phi}|0\rangle\langle 0| \,, \quad M_1 =
\tan\phi|0\rangle \langle 0| + |1\rangle\langle 1| \,,
\end{equation}
where $2\phi = \pi/2-\theta$.  For $T=0$, the effect of the
measurement is easily calculated by recalling that this state can be expressed as
$(\text{CZ}_{\mathcal{L}})|\theta\rangle^{\otimes\mathcal{L}}$. Both measurement operators $M_0$ and $M_1$ commute with $\text{CZ}_{\mathcal{L}}$, and thus we can consider their effect on
the state $|\theta\rangle$ at each site.  If the `0' outcome is
obtained, which occurs with probability $p_0=\sin\theta$, the
post-measurement state is $|0\rangle$; if `1' is obtained, the
resulting state is $|{+}\rangle$.  If the operation
$\text{CZ}_{\mathcal{L}}$ is applied to a lattice of qubits where a
subset $\mathcal{L}'$ (corresponding to those qubits for which the
measurement outcome `1' was obtained) has each qubit prepared in the $|{+}\rangle$ state and all remaining qubits prepared in $|0\rangle$, the result is an ideal cluster state on $\mathcal{L}'$ where the remaining sites remain unentangled. We will refer to this procedure as \emph{local filtering}.

Similar results hold for the thermal state, with $\cos2\phi =
\tanh(\beta\Delta/2)\sin\theta$.  The resulting state is a thermal
state of the $\theta=0$ Hamiltonian on $\mathcal{L}'$ (the subset of the lattice where the measurement results `1' were obtained) with an increased temperature $T'=(k_B\beta')^{-1}$, given by
\begin{equation}
  \tanh(\beta'\Delta/2) = \tfrac{\cos\theta}{\sqrt{1-\tanh^2(\beta\Delta/2)\sin^2\theta}} \tanh(\beta\Delta/2)\,.
\end{equation}
The probability that, for a given site, the measurement yields `1' is $p_1 = 1-\tanh(\beta\Delta/2)\sin\theta$.

With this filtering, we can argue the existence of a finite region of parameter space where MBQC is definitely possible on a cubic lattice using ideal single-qubit measurements.  Define $\overline{\mathcal{L}'}$ to be the subset of qubits for which the `0' outcome occurred, i.e., the complement of $\mathcal{L}'$.  For $\theta\ll 1$ the measurement disentangles the qubits in $\overline{\mathcal{L}'}$ from the rest of the cluster state, leaving them in the state $|0\rangle$. We then randomly flip each qubit in $\overline{\mathcal{L}'}$, such that they are described by the completely mixed state. We also apply a $Z$ gate to all neighbouring qubits to those that have been flipped into the state $|1\rangle$. The effect of this further processing is to prepare a $\theta=0$ thermal cluster state for which complete dephasing has been applied qubits in $\overline{\mathcal{L}'}$. We then discard the measurement record and the result is a thermal cluster state on the entire lattice $\mathcal{L}$ for which each qubit has been passed through an effective dephasing channel $\chi(\rho) =
(1-p)\rho + pZ\rho Z$, with $p = \tanh(\beta\Delta/2)\sin\theta$. This is just the same state as the $\theta=0$ thermal state with $T$ chosen such that $p=(1+\exp(\Delta/k_B T))^{-1}$.

Thus, a thermal $\theta\neq 0$ state can be converted by local
measurements into an ideal cluster state subjected to dephasing noise. With this fact, we use the results of~\cite{Rau06}, which demonstrate that MBQC can proceed using the cluster state on a body-centred cubic (BCC) lattice with dephasing noise up to $p_c \sim 2.9 \times 10^{-2}$. This bound defines a region at low $T$ and $\theta$ wherein every state is useful for MBQC given ideal measurements.  Fig.~\ref{fig:PhaseDiagram} shows this region for a cubic lattice (noting that a cubic lattice can be converted into a BCC lattice with local $Z$ measurements, which are unaffected by the noise), labeled as \emph{Region Q}.  At $\theta=0$, the boundary corresponds to a temperature of $T_c \sim 0.28 \Delta$.

However, these results may be too conservative.  Consider the $\theta\neq 0$ ground state, and apply the local filtering measurement.  Again, the effect of this measurement is to  disentangle some qubits from the lattice, and leave the remaining qubits in a cluster state.  One can then investigate whether the resulting subset $\mathcal{L}'$ contains a cluster of neighbouring sites that spans the lattice $\mathcal{L}$. This will occur with certainty if the success probability $1-\sin \theta$ is above the site percolation threshold, $p_c^{\text{site}}$. So if $1{-}\sin\theta > p_c^{\text{site}}$, then the resulting cluster state on $\mathcal{L}'$ is a universal resource for MBQC~\cite{Kie07,Bro08}. This lower bound (valid only at $T=0$) is much larger than the conservative lower bound on the critical value of $\theta$ obtained via the dephasing argument above.

A similar question can be asked of the thermal cluster state; however, the results of~\cite{Rau05,Rau06,Rau07} do not directly apply because it is not clear how to convert the thermal cluster state on the irregular set $\mathcal{L}'$ into the BCC lattice for which error thresholds are known. (A direct conversion using the methods of~\cite{Kie07,Bro08} would require performing local $X$ measurements, which do not commute with the noise.) As a result it is not clear that MBQC can in fact be performed. Nevertheless,
Fig.~\ref{fig:PhaseDiagram} shows the region Q' for which the success probability of the filtering operation is greater than the site percolation threshold $1{-}\sin\theta > p_c^{\text{site}}$, with a resulting temperature $T'$ less than the critical temperature for universal MBQC.  An alternative approach would be to apply the error correction procedure of~\cite{Stace09} which corrects the effects of both finite temperature and qubit loss errors.  The performance of this scheme in a three dimensional cluster is currently unknown, however it may yield better performance than the  strategies described above in some regions of parameter space.

\section{Conclusions}
\label{sec:Conclusions}

We have shown that the thermal equilibrium
states of Eq.~(\ref{eq:ClusterLocalZ}) undergo a transition in their usefulness for MBQC, from a region of parameter space where every
state is a universal resource, to one where every state is
efficiently classically simulatable. In spite of this dramatic
change in computational power, these states do not exhibit any
corresponding phase transition.

We conclude by contrasting our results with some related work.  First, in a similar model with a transverse rather than longitudinal local field, the system \emph{does} exhibit a zero-temperature phase transition at sufficiently high field strength, and correlation
functions can be identified which characterize a phase for which a set of MBQC gates can be performed over arbitrary ranges~\cite{Doh09}.  Thus, a transition in computational power coincides with a quantum phase transition in this model.

Second, non-analytic behaviour in long-range entanglement quantities such as localizable entanglement~\cite{Pop05} does not necessarily indicate a phase transition. In our model, for $\sin\theta < 1{-}p_c^{\text{site}}$, the localizable entanglement in the ground state is precisely equal to 1 ebit at all length scales, and at $\theta=\pi/2$ it is equal to zero (because the state is a product state).  There is no analytic function fulfilling these requirements, and so the localizable entanglement is a non-analytic function of $\theta$, indicating a sharp transition somewhere in the interval $1{-}p_c^{\text{site}} < \sin\theta < 1$.  Again, we emphasize that there is no quantum phase transition at any value of $\theta$ in this model, and so we have an example of a system where a non-analyticity in the localizable entanglement length does \emph{not} identify an underlying quantum phase transition.  This in contrast to the situation for a large number of models discussed in the literature~\cite{Pop05} including the cluster Hamiltonian with a local $X$ field~\cite{Pac04,Doh09}.

\begin{acknowledgments}
We thank Norbert Schuch and Karl Gerd Vollbrecht for highlighting the importance of the site projection measurement probabilities, and outlining the method presented in Appendix~\ref{app:Sampling} to incorporate these projections into our classical simulation.  We thank Damian Abasto for helpful comments and corrections.  S. D. Barrett acknowledges the support of the EPSRC.  S. D. Bartlett and ACD acknowledge the support of the Australian Research Council. TR acknowledges the support of the EPSRC and the QIP-IRC.
\end{acknowledgments}

\appendix

\section{The sampling algorithm} \label{app:Sampling}

The sampling of PEPS bonds to generate a single pure state instance
of the thermal state requires some care. We must sample from the
ensemble of bond instances \emph{after} each of the site projections
$A = |0\rangle\langle0 \dots 0| + |1 \rangle\langle 1 \dots 1|$.  To
obtain the correct distribution, the projections $A$ should be
viewed as physical measurement operations, and one should sample
from the posterior distribution conditional on the success of the
operations. However, the success of the projection at a site can
vary depending on the choice of pure bond states at the site. For
example, a site containing a virtual qubit in the state $|0\rangle$,
and another virtual qubit in $|1\rangle$, will yield zero
conditional probability for the projection, while a site with all
qubits in the pure state $|0\rangle$ will guarantee success for the
projection. Consequently, we cannot sample directly from the
ensemble of pure bond states according to the probabilities in the
decomposition of Eq.~(\ref{eqn:BSA}); instead, we use a sampling procedure where each step is conditioned on the success of the site projections. Here we describe an algorithm that samples efficiently from the appropriate distribution, and show below that the algorithm samples the correct distribution for the whole system.  The key
requirement for this simulation to be efficient is that the success of the measurement $A_i$ at site $i$ is independent of the other sites in the lattice.

The following algorithm efficiently samples from the posterior bond
distribution.  That is, it reproduces the distribution
$p\left(\left\{\rho^{\{ij\}} \right\} | A_1\otimes A_2 \otimes
\ldots \otimes A_N \right)$ of a particular configuration
$\{\rho^{\{ij\}}\}$ of all bonds in the lattice conditional on
success of each of the projectors $A_i$.
\vspace{1em}
\algsetup{indent=0.5em}
\begin{algorithmic}
\STATE set all virtual qubits to \emph{empty}
\FORALL{ bonds $(i,j)$}
\STATE set any \emph{empty} qubits in $\{i.*\setminus i.j\}$ and $\{j.*\setminus j.i\}$ to the state $\rho_0 = \mathrm{tr}_2[\rho_{\mathrm{bond}}]$
\STATE sample from the distribution
\begin{displaymath}
p[\rho^{\{i,j\}}|A_i \otimes A_j, \rho^{\{i.*\setminus i.j\}}, \rho^{\{j.*\setminus j.i\}}]
\end{displaymath}
\STATE set the state of qubits $i.j$ and $j.i$ to the corresponding bond state
\ENDFOR
\end{algorithmic}
\vspace{1em}
Here, $\{i.*\setminus i.j\}$ is the set of virtual qubits at site
$i$ except for the one associated with the bond $(i,j)$, and
$p[\rho^{\{i,j\}}|A_i \otimes A_j, \rho^{\{i.*\setminus i.j\}},
\rho^{\{j.*\setminus j.i\}}]$ is the posterior probability for the
bond $(i,j)$, given the state of the other virtual qubits at sites
$i$ and $j$ and given that the projections $A_i$ and $A_j$ succeed.
We can calculate this distribution straightforwardly from Bayes'
rule.

%\section{Validity of the algorithm}

% Give the final result.

We now show that the above algorithm indeed samples from the correct probability distribution, which is the posterior distribution for
the bond configuration, conditioned on the success of all $N$
projections. The desired distribution may be written, using Bayes'
rule, as
\begin{widetext}
\begin{equation} \label{PosteriorWholeState}
p\left(\left\{\rho^{\{ij\}} \right\}  |  A_1\otimes A_2  \otimes \ldots \otimes  A_N \right) =
\frac{p\left( A_1\otimes A_2  \ldots \otimes  A_N |\left\{\rho^{\{ij\}} \right\}   \right) \times p\left(\left\{\rho^{\{ij\}} \right\}   \right)}
{p\left( A_1\otimes A_2  \otimes \ldots \otimes  A_N \right)} \,.
\end{equation}
%
% Give the bond-by-bond expression that the algorithm samples from.
On the other hand, the algorithm described above samples, on a bond-by-bond basis, from the distribution
\begin{align}
p_{\mathrm{alg}}\left(\left\{\rho^{\{ij\}} \right\}  \right) & = \prod_{\{ij\}} p[\rho^{\{ij\}}|A_i \otimes A_j, \rho^{\{i.*\setminus i.j\}}, \rho^{\{j.*\setminus j.i\}}] \nonumber \\
& =  \prod_{\{ij\}}  \left. \frac{p[  A_i \otimes A_j | \rho^{\{ij\}} , \rho^{\{i.*\setminus i.j\}}, \rho^{\{j.*\setminus j.i\}}] \times p^{\{ij\}}}
{p[  A_i \otimes A_j | \rho^{\{i.*\setminus i.j\}}, \rho^{\{j.*\setminus j.i\}}]} \right .  \label{PosteriorAlgorithm} \,.
\end{align}
Here, the product runs over all bonds $\{ij\}$ in the lattice, $p[
A_i \otimes A_j | \rho^{\{ij\}} , \rho^{\{i.*\setminus i.j\}},
\rho^{\{j.*\setminus j.i\}}] $ is the probability that the
projections at site $i$ and $j$ succeed, conditioned on the bond
$\{ij\}$ being in state $\rho^{\{ij\}}$, and $p^{\{ij\}}$ is the
prior distribution for each bond, as given by the probabilities
$\{p_e, p_{s,\mu}\}$ of Eq.~(\ref{eqn:BSA}). The denominator ${p[
A_i \otimes A_j | \rho^{\{i.*\setminus i.j\}}, \rho^{\{j.*\setminus
j.i\}}]}$ denotes the total probability that the projections at site
$i$ and $j$ succeed, conditioned only on the state of the virtual
qubits ${\{i.*\setminus i.j\}}$ and ${\{j.*\setminus j.i\}}$.

% Factorisation argument.
In order to show the equivalence of the two distributions of Eq. (\ref{PosteriorWholeState}) and Eq. (\ref{PosteriorAlgorithm}), first consider the denominators in  Eq. (\ref{PosteriorAlgorithm}). These may be written as
\begin{equation}
p[  A_i \otimes A_j | \rho^{\{i.*\setminus i.j\}}, \rho^{\{j.*\setminus j.i\}}]  = \sum_k {p[  A_i \otimes A_j | \rho_k^{\{ij\}} , \rho^{\{i.*\setminus i.j\}}, \rho^{\{j.*\setminus j.i\}}] \times p_k^{\{ij\}}} \,,
\end{equation}
where $\rho_k^{\{ij\}}$ are the elements of the ensemble decomposition of Eq.~(\ref{eqn:BSA}), and $p_k^{\{ij\}}$ the corresponding weights. This expression may be written
\begin{align}
p[  A_i \otimes A_j | \rho^{\{i.*\setminus i.j\}}, \rho^{\{j.*\setminus j.i\}}] & =\sum_k \mathrm{tr}[(A_i \otimes A_j)  \rho^{\{i.*\setminus i.j\}} \otimes \rho_k^{\{ij\}} \otimes \rho^{\{j.*\setminus j.i\}} (A_i \otimes A_j)^\dag]    p_k^{\{ij\}} \nonumber \\
& =  \mathrm{tr}[(A_i \otimes A_j )  \rho^{\{i.*\setminus i.j\}} \otimes \rho_{\mathrm{bond}}^{\{ij\}} \otimes \rho^{\{j.*\setminus j.i\}} (A_i \otimes A_j)^\dag] \,, \label{traceAA}
\end{align}
where $\rho_{\mathrm{bond}}^{\{ij\}}$ denotes the total state of the
bond. In Sec.~\ref{sec:AppB}, we show that for $\theta \ge 0$, the
thermal bond state is of the form $\rho_{\mathrm{bond}}
=\tfrac{1}{4} (I + \alpha X \otimes Z + \gamma Z \otimes I) (I +
\alpha Z \otimes X + \gamma I\otimes Z)$, where $\alpha$ and $\gamma$
are parameters, determined numerically for a given temperature and
field strength (see Sec.~\ref{sec:AppB}). For \emph{any} physical
values of $\alpha$, $\gamma$,  $\rho^{\{i.*\setminus i.j\}}$ and
$\rho^{\{j.*\setminus j.i\}}$, one finds that this expression may be
written
\begin{equation} \label{DenominatorFinal}
p[  A_i \otimes A_j | \rho^{\{i.*\setminus i.j\}}, \rho^{\{j.*\setminus j.i\}}]  =   p[A_i | \rho^{\{i.*\setminus i.j\}} , \rho_0^{\{i.j\}} ] \times p[A_j |\rho^{\{j.*\setminus j.i\}},\rho_0^{\{j.i\}} ] \,.
\end{equation}

%       Then numerator
The numerators in Eq. (\ref{PosteriorAlgorithm}) contain likelihood factors of the form
\begin{equation}
p[  A_i \otimes A_j | \rho^{\{ij\}} , \rho^{\{i.*\setminus i.j\}}, \rho^{\{j.*\setminus j.i\}}]  =  \mathrm{tr}[(A_i \otimes A_j )  \rho^{\{i.*\setminus i.j\}} \otimes \rho^{\{ij\}} \otimes \rho^{\{j.*\setminus j.i\}} (A_i \otimes A_j)^\dag] \,.
\end{equation}
If $ \rho^{\{ij\}} =  \rho^{\{i.j\}} \otimes  \rho^{\{j.i\}}$ then it is clear that this expression can be written as a product of two probabilities (corresponding to independent outcomes $A_i$ and $A_j$ at sites $i$ and $j$ respectively). Conversely, as we show in the final section of this appendix, the entangled component of the decomposition is also of the form $\rho_e =\tfrac{1}{4} (I + \alpha_0 X \otimes Z +
\gamma_0 Z \otimes I) (I + \alpha_0 Z \otimes X + \gamma_0 I\otimes Z)$, where $\alpha_0$ and $\gamma_0$ are real parameters. Since, in this case, $ \rho^{\{ij\}} $ is of the same form as the state $\rho_{\mathrm{bond}}^{\{ij\}}$ appearing in Eq. (\ref{traceAA}), it follows from Eq.(\ref{DenominatorFinal}) that the expression can also be written as a product of two probabilities. Thus this term may be written as the product:
\begin{equation}  \label{LikelihoodFinal}
p[  A_i \otimes A_j | \rho^{\{ij\}} , \rho^{\{i.*\setminus i.j\}}, \rho^{\{j.*\setminus j.i\}}]  =p[A_i | \rho^{\{i.j\}}, \rho^{\{i.*\setminus i.j\}} ] \times p[ A_j | \rho^{\{j.i\}},  \rho^{\{j.*\setminus j.i\}} ]   \,.
\end{equation}

Note that generalizations of Eq. (\ref{DenominatorFinal}) and Eq. (\ref{LikelihoodFinal}) to arbitrary numbers of sites also hold. For example, in the three site case, where site $j$ neighbours both site $i$ and site $k$, we find
\begin{multline} \label{DenominatorThreeSites}
p[  A_i \otimes A_j \otimes A_k | \rho^{\{i.*\setminus i.j\}},
\rho^{\{j.*\setminus \{j.i,j.k\}\}}, \rho^{\{k.*\setminus k.j\}}] \\
= p[A_i | \rho^{\{i.*\setminus i.j\}} , \rho_0^{\{i.j\}} ] \times
p[A_j |\rho^{\{j.*\setminus \{j.i,j.k\}\}},
\rho_0^{\{j.i\}},\rho_0^{\{j.k\}} ] \times p[A_k |
\rho^{\{k.*\setminus k.j\}} , \rho_0^{\{k.j\}} ] \,,
\end{multline}
and also
\begin{multline} \label{LikelihoodThreeSites}
p[  A_i \otimes A_j \otimes A_k |  \rho^{\{ij\}} ,  \rho^{\{jk\}},
\rho^{\{i.*\setminus i.j\}}, \rho^{\{j.*\setminus \{j.i,j.k\}\}},
\rho^{\{k.*\setminus k.j\}}] \\
 = p[A_i | \rho^{\{i.j\}} ,
\rho^{\{i.*\setminus i.j\}}  ] \times p[A_j |
\rho^{\{j.i\}},\rho^{\{j.k\}}, \rho^{\{j.*\setminus \{j.i,j.k\}\}} ]
\times p[A_k | \rho^{\{k.j\}} , \rho^{\{k.*\setminus k.j\}} ] \,.
\end{multline}
One way to verify these expressions, and their generalizations to any number of sites, is as follows. Recall that the site projections of the PEPS representation commute with $\text{CZ}_{\mathcal{L}}$ gates applied on each of the bonds. Consider a single instance of the thermal ensemble of entangled and product state bonds, as shown in Figure~\ref{fig:PEPSpercolation}(b). Consider performing a $\text{CZ}_{\mathcal{L}}$ operation on every bond that is occupied by an entangled state, leaving the bonds with product states alone. The site projections are unaffected by this operation, however the entangled bond states have become completely unentangled and the resulting state is a product state between all sites. The success probabilities at each site are now clearly independent.

% Cancellation argument.
Using the factorized expressions Eq. (\ref{DenominatorFinal}) and Eq. (\ref{LikelihoodFinal}), Eq. (\ref{PosteriorAlgorithm}) becomes
\begin{align}
p_{\mathrm{alg}}  & \left(  \left\{  \rho^{\{ij\}} \right\}  \right)
=    \prod_{\{ij\}}  \left. \frac{p[A_i | \rho^{\{i.j\}},
\rho^{\{i.*\setminus i.j\}} ] \times p[ A_j | \rho^{\{j.i\}},
\rho^{\{j.*\setminus j.i\}} ] \times p^{\{ij\}}}
{p[A_i | \rho^{\{i.*\setminus i.j\}} , \rho_0^{\{i.j\}} ] \times p[A_j |\rho^{\{j.*\setminus j.i\}},\rho_0^{\{j.i\}} ]} \right . \nonumber \\
 & =  \prod_i \frac{p[A_i | \rho^{\{i.j_1\}}, \rho_0^{\{i.j_2\}}, \ldots, \rho_0^{\{i.j_d\}} ]
\times
p[A_i | \rho^{\{i.j_1\}}, \rho^{\{i.j_2\}}, \rho_0^{\{i.j_3\}} \ldots, \rho_0^{\{i.j_d\}} ]
\times \ldots \times
p[A_i | \rho^{\{i.j_1\}}, \rho^{\{i.j_2\}}, \ldots, \rho^{\{i.j_d\}} ]
}
{p[A_i | \rho_0^{\{i.j_1\}}, \rho_0^{\{i.j_2\}}, \ldots, \rho_0^{\{i.j_d\}} ]
\times
p[A_i | \rho^{\{i.j_1\}}, \rho_0^{\{i.j_2\}}, \ldots, \rho_0^{\{i.j_d\}} ]
\times \ldots \times
p[A_i | \rho^{\{i.j_1\}}, \rho^{\{i.j_2\}}, \ldots, \rho_0^{\{i.j_d\}} ]}
\nonumber \\
& \qquad \times \prod_{\{ij\}}   p^{\{ij\}} \label{ComplicatedFactorisedExpression}
\end{align}
where in the second line we have re-arranged the product such that
each factor now corresponds to a particular site of the lattice,
rather than a bond. The ordering of the virtual qubits at site $i$
implied by the indices $j_1 \ldots j_d$ corresponds to the order in
which the corresponding bonds are sampled in the algorithm.
Eq.~(\ref{ComplicatedFactorisedExpression}) can be simplified
considerably by noting that many repeated terms appear in both the
numerator and denominator, yielding
\begin{equation}
p_{\mathrm{alg}}   \left(  \left\{  \rho^{\{ij\}} \right\}  \right) =  \frac{ \prod_i p[A_i | \rho^{\{i.j_1\}}, \rho^{\{i.j_2\}}, \ldots, \rho^{\{i.j_d\}} ] \times \prod_{\{ij\}}   p^{\{ij\}}}
{ \prod_i p[A_i | \rho_0^{\{i.j_1\}}, \rho_0^{\{i.j_2\}}, \ldots, \rho_0^{\{i.j_d\}} ]}
 \,. \label{SimpleFactorisedExpression}
\end{equation}
\end{widetext}

% Finally show by inverse factorisation that we get back to our first equation.
Finally, by observing that the expressions $p\left( A_1\otimes A_2  \ldots \otimes  A_N |\left\{\rho^{\{ij\}} \right\}   \right)$ and $p\left( A_1\otimes A_2  \otimes \ldots \otimes  A_N \right)$ of Eq. (\ref{PosteriorWholeState}) may be written in a factorized form (by making use of the appropriate generalization of Eqs. (\ref{DenominatorThreeSites}) and (\ref{LikelihoodThreeSites}), we can identify each term in Eq. (\ref{SimpleFactorisedExpression}) with the corresponding term in Eq. (\ref{PosteriorWholeState}). Thus, our algorithm indeed samples from the correct distribution for the whole state.

\section{Thermal states with non-zero local field} \label{sec:AppB}

In the non-zero field case the thermal bonds are determined by the
requirement that they project, under $A_i$ at each site, to the
correct thermal cluster state.

In the zero temperature case it is possible to obtain an analytical
expression for this bond state. It is straightforward to see that
the states
\begin{multline}
\rho_{\mathrm{bond}}(T=0, \theta) \\ = \frac{1}{4} ( I + X\otimes Z
\cos 2 \phi - Z \otimes I \sin 2 \phi ) \\ \times ( I + Z \otimes X \cos 2
\phi - I \otimes Z \sin 2 \phi )\,,
\end{multline}
yield the correct pure cluster state, provided that $\tan ^d (\phi + \pi/4) = \tan (\theta/2 + \pi/4)$, where $d$ is the coordination
number of the lattice. However, for finite temperatures it is slightly more difficult to obtain the relations between the bond parameters and the values of $T$ and $\theta$. As stated in the main text, the bond states take the general form
\begin{equation}
  \rho_{\mathrm{bond}}(T, \theta) = \frac{1}{4} ( I + \alpha    X\otimes
  Z  + \gamma Z \otimes I  ) ( I + \alpha Z \otimes X  + \gamma I
\otimes Z)\,,
\end{equation}
where the parameters $\alpha$ and $\gamma$ obey
\begin{align}
  \alpha^d &= \tanh \frac{\beta \Delta }{2} \cos \theta \sum
_{\mathrm{j \hspace{1mm} even}}
{d \choose j } \gamma^j \,, \nonumber \\
\sum _{\mathrm{j\hspace{1mm} odd}} {d \choose j } \gamma^j &=- \tanh
\frac{\beta \Delta }{2} \sin \theta \sum _{\mathrm{j \hspace{1mm}
even}} {d \choose j } \gamma^j \,,
\end{align}
while also being constrained to producing physical bond states. In one dimension these equations are relatively straightforward and may be solved without much trouble, however for the cubic lattice the conditions involve sixth order polynomials in $\gamma$ and so we opt to solve these constraints numerically, and find that for any $T$ or $0\le \theta < \pi/2$ we obtain an appropriate thermal bond state.

To calculate the classical simulation bound from percolation on the lattice, the thermal bond state must be decomposed into an  entangled part and a separable part. Instead of calculating the Best Separable Approximation (BSA) for the thermal state, we use the state $\rho_e = \rho_{\mathrm{bond}} (T=0, \theta)$ as the (non-maximally) entangled pure bond state in the decomposition $\rho_{\mathrm{bond}} = p_e \rho_e + (1-p_e) \rho_s$ and choose the largest $p_e$ for which $\rho_s$ remains separable.

%\newpage

\end{document}